\documentclass[twocolumn,iop,revtex4]{openjournal}
\usepackage[utf8]{inputenc}
\usepackage{times}
\usepackage{natbib} 
\usepackage{epsfig} 
\usepackage{graphicx} 
\usepackage[dvipsnames]{xcolor}
\usepackage{xspace}
\usepackage{aas_macros} 
\usepackage{amssymb}
\usepackage{amsmath}
\usepackage[title]{appendix}
\usepackage{hyperref}	% Hyperlinks
\hypersetup{colorlinks=true,linkcolor=blue,citecolor=blue,filecolor=blue,urlcolor=blue}
\usepackage[caption=false]{subfig}
\usepackage{soul}                          % provides \hl{} for highlighting
\usepackage{ulem} 
\usepackage{xifthen}

\newcommand{\kms} {km $\rm{s^{-1}}$}

\newcommand{\change}[2][]{%
\ifthenelse{\isempty{#2}}{{\color{ForestGreen}{#1}}}%
{{\color{RedOrange}\sout{#1}}{\color{ForestGreen}{#2}}}%
}

\newcommand*{\Euclid}{\textit{Euclid}}

\newcommand*{\AckInstitutions}{a number of agencies and
  institutes that have supported the development of \Euclid, in
  particular
the Academy of Finland, the Agenzia Spaziale Italiana,
the Belgian Science Policy, the Canadian Euclid Consortium, the Centre
National d'Etudes Spatiales, the Deutsches Zentrum f\"ur Luft- und
Raumfahrt, the Danish Space Research Institute, the Funda\c{c}\~{a}o
para a Ci\^{e}ncia e a Tecnologia, the Ministerio de Economia y
Competitividad, the National Aeronautics and Space Administration, the
Netherlandse Onderzoekschool Voor Astronomie, the Norwegian Space
Agency, the Romanian Space Agency, the State Secretariat for
Education, Research and Innovation (SERI) at the Swiss Space Office
(SSO), and the United Kingdom Space Agency. A complete and detailed
  list is available on the \Euclid\ web site 
(\texttt{http://www.euclid-ec.org}).}

\newcommand{\AckECol}{The authors acknowledge the Euclid
  Collaboration, the European Space Agency, and \AckInstitutions}

\begin{document}

 \title{\emph{Euclid}: Forecasts for $k$-cut $3 \times 2$ Point Statistics$^{\star}$} \thanks{$^\star$ This paper is published on behalf of the Euclid Consortium.}

%% please do not edit the author list -- contact ECEB Bureau for changes
\author{P.L.~Taylor$^{1}$\thanks{\email{peter.taylor@jpl.nasa.gov}}, T.~Kitching$^{2}$, V.F.~Cardone$^{3}$, A.~Fert\'e$^{1}$, E.M.~Huff$^{1}$, F.~Bernardeau$^{4,5}$, J.~Rhodes$^{1}$, A.C.~Deshpande$^{2}$, I.~Tutusaus$^{6,7}$, A.~Pourtsidou$^{8}$, S.~Camera$^{9,10}$, C.~Carbone$^{11}$, S.~Casas$^{12}$, M.~Martinelli$^{13}$, V.~Pettorino$^{12}$, Z.~Sakr$^{14,15}$, D.~Sapone$^{16}$, V.~Yankelevich$^{17}$, N.~Auricchio$^{18}$, A.~Balestra$^{19}$, C.~Bodendorf$^{20}$, D.~Bonino$^{21}$, A.~Boucaud$^{22}$, E.~Branchini$^{3,23,24}$, M.~Brescia$^{25}$, V.~Capobianco$^{21}$, J.~Carretero$^{26}$, M.~Castellano$^{3}$, S.~Cavuoti$^{25,27,28}$, A.~Cimatti$^{29,30}$, R.~Cledassou$^{31}$, G.~Congedo$^{32}$, L.~Conversi$^{33,34}$, L.~Corcione$^{21}$, M.~Cropper$^{2}$, E.~Franceschi$^{18}$, B.~Garilli$^{11}$, B.~Gillis$^{32}$, C.~Giocoli$^{18,35}$, L.~Guzzo$^{36,37}$, S.V.H.~Haugan$^{38}$, W.~Holmes$^{1}$, F.~Hormuth$^{39}$, K.~Jahnke$^{40}$, S.~Kermiche$^{41}$, M.~Kilbinger$^{12}$, M.~Kunz$^{42}$, H.~Kurki-Suonio$^{43}$, S.~Ligori$^{21}$, P.~B.~Lilje$^{38}$, I.~Lloro$^{44}$, O.~Marggraf$^{45}$, K.~Markovic$^{1}$, R.~Massey$^{46}$, E.~Medinaceli$^{47}$, S.~Mei$^{48}$, M.~Meneghetti$^{18,35,49}$, G.~Meylan$^{50}$, M.~Moresco$^{18,30}$, B.~Morin$^{12}$, L.~Moscardini$^{18,30,35}$, S.~Niemi$^{51}$, C.~Padilla$^{26}$, S.~Paltani$^{52}$, F.~Pasian$^{53}$, K.~Pedersen$^{54}$, W.J.~Percival$^{55,56,57}$, S.~Pires$^{12}$, G.~Polenta$^{58}$, M.~Poncet$^{31}$, L.~Popa$^{59}$, F.~Raison$^{20}$, M.~Roncarelli$^{18,30}$, E.~Rossetti$^{30}$, R.~Saglia$^{20,60}$, P.~Schneider$^{45}$, A.~Secroun$^{41}$, G.~Seidel$^{40}$, S.~Serrano$^{6,7}$, C.~Sirignano$^{61,62}$, G.~Sirri$^{35}$, F.~Sureau$^{12}$, P.~Tallada Cresp\'{i}$^{63}$, D.~Tavagnacco$^{53}$, A.N.~Taylor$^{32}$, H.I.~Teplitz$^{49,64}$, I.~Tereno$^{65,66}$, R.~Toledo-Moreo$^{67}$, E.A.~Valentijn$^{68}$, L.~Valenziano$^{18,35}$, T.~Vassallo$^{60}$, Y.~Wang$^{64}$, J.~Weller$^{20,60}$, A.~Zacchei$^{53}$, J.~Zoubian$^{41}$}
   
\affiliation{(Affiliations can be found after the references)}
\footnote{\textcopyright 2021. All rights reserved.}

\begin{abstract}
   Modelling uncertainties at small scales, i.e. high $k$ in the power spectrum $P(k)$, due to baryonic feedback, nonlinear structure growth and the fact that galaxies are biased tracers poses a significant obstacle to fully leverage the constraining power of the {\it Euclid} wide-field survey. $k$-cut cosmic shear has recently been proposed as a method to optimally remove sensitivity to these scales while preserving usable information.
   In this paper we generalise the $k$-cut cosmic shear formalism to $3 \times 2$ point statistics and estimate the loss of information for different $k$-cuts in a $3 \times 2$ point analysis of the {\it Euclid} data.
   Extending the Fisher matrix analysis of~\citet{blanchard2019euclid}, we assess the degradation in constraining power for different $k$-cuts. We work in the idealised case and assume the galaxy bias is linear, the covariance is Gaussian, while neglecting uncertainties due to photo-z errors and baryonic feedback.
   We find that taking a $k$-cut at $2.6 \ h \ {\rm Mpc} ^{-1}$ yields a dark energy Figure of Merit (FOM) of 1018. This is comparable to taking a weak lensing cut at $\ell = 5000$ and a galaxy clustering and galaxy-galaxy lensing cut at $\ell = 3000$ in a traditional $3 \times 2$ point analysis. We also find that the fraction of the observed galaxies used in the photometric clustering part of the analysis is one of the main drivers of the FOM. Removing $50 \%  \ (90 \%)$ of the clustering galaxies decreases the FOM by $19 \% \ (62 \%)$.
Given that the FOM depends so heavily on the fraction of galaxies used in the clustering analysis, extensive efforts should be made to handle the real-world systematics present when extending the analysis beyond the luminous red galaxy (LRG) sample.

\end{abstract}


\section{Introduction} \label{sec:Introduction}

The {\it Euclid}\footnote{\url{http://euclid-ec.org}} wide-field survey will measure the shapes and photometric redshifts of approximately $1.5$ billion galaxies out to redshifts $z\sim 2$ \citep{laureijs2010euclid}. Cosmic shear, photometric clustering, and the correlation between background `source galaxies' and foreground `lens galaxies' -- referred to as galaxy-galaxy lensing -- will help constrain both the growth of structure and the background expansion of the late Universe. The galaxy-galaxy lensing signal is particularly important for constraining nuisance parameters which are marginalised over, to avoid a large degradation in constraining power~\citep{tutusaus2020euclid}. At the two-point level these three signals are referred to as $3 \times 2$ point statistics.  
\par Compared to today's photometric surveys, the {\it Euclid} wide-field survey offers massive increases in statistical constraining power; hence $3 \times 2$ point analyses risk becoming limited by systematic effects. Modelling uncertainties at small scales is one of the primary causes as non-linear structure growth, baryonic feedback~\citep{semboloni2011quantifying}, intrinsic alignment (IA) of galaxies\footnote{Since the IA kernels are different from the lensing efficiency kernels, the $k$-cut developed in this work does not fully alleviate small-scale IA modelling bias.}~\citep{kiessling2015galaxy}, and galaxy bias~\citep{desjacques2018large} are all uncertain at small scales.
\par Broadly speaking there are two ways to tackle these uncertainties. One can attempt to model the small scales -- potentially including a few free parameters that are either marginalised over in a likelihood analysis or calibrated against simulations -- or scales can be cut. The two approaches are typically hybridised. For example, recent studies of Hyper-Suprime Cam (HSC), Dark Energy Survey (DES), and Kilo-Degree Survey (KiDS) data sets~\citep{hikage2018cosmology,troxel2018dark, asgari2020kids}, all marginalised over IA parameters while cutting small angular scales. 
\par The objective should always be to model small scales accurately. However, if scales must be cut to mitigate model bias, it is important that a maximal amount of `useful' information at large scales is retained. Removing principal components where there is large disagreement between models (PCA) is a possible approach~\citep{eifler2015accounting, huang2019modelling, huang2020dark}. However in many circumstances it is known {\it a priori} that small scales are the most severely affected, so it is simpler and more physical to just cut these directly. Unlike PCA there is no requirement to have multiple competing models and no need to repeat the procedure for each systematic effect.
\par Most $3 \times 2$ point analyses take na\"ive angular scale or inverse angular scale cuts (i.e. $\ell$-cuts in harmonic space, $\theta$-cuts in configuration space or more optimally discrete modes~\citep{asgari2020kids+} when using Complete Orthogonal Sets of E/B-Integrals, abbreviated COSEBIs). None of these correspond exactly to cutting small physical scales. In this paper we present $k$-cut $3\times 2$ point statistics, which are constructed to optimally filter out small scales.\footnote{We choose to work in harmonic space for the remainder of the paper, but the arguments are readily generalisable to configuration space as in~\citet{taylor2020x}.} The objective of this work is to demonstrate how this formalism could be used in {\it Euclid} to remove sensitivity to small uncertain scales and provide forecasts for different scale cuts.
\par We note from the small angle approximation (or alternatively the Limber relation) that for structure at a comoving distance $\chi$ we have $\ell \sim k \chi$, so that each $\ell$-mode corresponds to a unique inverse physical scale, $k$. Thus, in the galaxy clustering case, cutting all $\ell > k \chi $ after defining a `typical' distance $\chi$ to each narrow tomographic bin~\citep{lanusse20153d} removes sensitivity to small scales (modes larger than $k$ in the matter power spectrum).
\par This argument is not as straightforward for cosmic shear and galaxy-galaxy lensing because the lensing efficiency kernels are broad, so the lensing signal of galaxies inside a very narrow tomographic bin are sensitive to structure over a broad range in redshift. To overcome this issue, one can apply the Bernardeau-Nishimichi-Taruya (BNT) transformation~\citep{bernardeau2014cosmic}. This is a linear combination of tomographic bins which results in a set of kernels that are narrow in redshift. Then one can take tomographic bin-dependent $\ell$-cuts to remove sensitivity to small scales. This is known as $k$-cut cosmic shear~\citep{taylor2018k-cut} in harmonic space and $x$-cut cosmic shear~\citep{taylor2020x} in configuration space (\citealt{huterer2005nulling} proposed a similar nulling scheme). Simultaneously taking a bin-dependent angular scale cut for the galaxy-clustering auto-spectra~\citep{lanusse20153d} defines a $3 \times 2$ point statistic which is insensitive to small scale information. We refer to these as $k$-cut $3 \times 2$ point statistics.
\par While it is important to remove small scales which are not accurately modelled, this is not the only cut made in $3 \times 2$ point analyses. On the galaxy clustering side, it is typical to perform the analysis on a sub-population of the observed galaxies (or an external clustering data set). For example the Kilo-Degree Survey (KiDS-1000) $3 \times 2$ point analysis~\citep{heymans2020kids} did not use the photometric data for the clustering part of the analysis, and instead  used external spectroscopic data from the Baryon Oscillation Spectroscopic Survey (BOSS)~\citep{ross2020completed} and 2-degree Field Lensing Survey (2dFLenS)~\citep{blake20162}. Meanwhile the DES year 1 (DESY1) analysis~\citep{abbott2018dark, elvin2018dark} took only luminous red galaxies (LRGs) using the {\it red-sequence matched-filter galaxy catalog algorithm} ({\tt REDMAGIC})~\citep{rozo2016redmagic}.
In total $26$ million `source' galaxies were used in the DESY1 analysis, while only $650\,000$ `lens' galaxies were used in the clustering analysis. This amounts to approximately $2.5 \%$ of the available galaxies. 
\par LRGs make ideal targets since they are bright, making selection effects less important, and there exists a tight photometric colour-redshift relation~\citep{rozo2016redmagic}. To expand beyond the typical LRG sample would require careful calibration of photometric redshifts, a sufficiently flexible galaxy bias model, $b(k,z)$, to handle the expanded multiple tracer population~\citep{kauffmann1997galaxy} and thorough mitigation of selection effects~\citep{elvin2018dark} including blending, which will become more important for fainter galaxies.
\par In this paper we do not attempt to answer the question of how the lens galaxy sample should be extended beyond the LRG subsample. Rather we examine the trade-off between taking a larger $k$-cut and including a larger fraction of the available lens galaxies in the clustering analysis. 
\par  The structure of this paper is as follows. In Sect.~\ref{sec:Formalism} we present the $k$-cut $3 \times 2$ point statistics and review the Fisher matrix formalism. We present the results in Sect.~\ref{sec:Results} before concluding in Sect.~\ref{sec:Conslusions}.

\section{Formalism} \label{sec:Formalism}

\begin{table}[hbt!]
\caption{The fiducial parameters and survey set-up used in this paper are from~\citet{blanchard2019euclid} (EF19) assuming a spatially flat cosmology. We refer the reader to this work for detailed overview of the modelling assumptions. We also indicate which cosmological and nuisance parameters are fixed; all other parameters are varied in the Fisher analysis.}
\label{table:params}
\begin{center}
\begin{ruledtabular}
\begin{tabular}{ lccccccc }
  Parameter &  Value   \\
  \hline
  \hline
  Survey Area $[{\rm deg} ^2]$ & $15 \ 000$ \\
  Number of Galaxies $[{\rm arcmin}^{-2}]$ & 30 \\
  $\sigma_\epsilon$ & 0.3 \\
  Number of Tomographic Bins & 10\\
  $[z_{\rm min}, z_{\rm max}]$ & $[0.0, 2.5] $\footnote{Redshift limits before photometric smoothing.}\\
  \hline
  $\sigma_8$ & 0.816  \\
  $\Omega_{\rm m}$ 	&  0.32  \\
  $\Omega_{\rm b}$ 	& 0.05   \\
  $\sum m_\nu$ $[{\rm eV}]$  & 0.06 (fixed)\\
  $h_0$ 		&  0.67  \\
  $n_{\rm s}$ 		& 0.96  \\
  $w_0$ 		& $-1.0$ \\
  $w_a$ 		& 0.0 \\
  \hline
  $A_{\rm IA}$  &  1.72 \\
  $C_{\rm IA}$  &  0.0134 (fixed) \\
  $\eta_{\rm IA}$ & $-0.41$  \\
  $\beta_{\rm IA}$ & $-2.17$  \\
  \hline
  $b_i$ \ {\rm for} \ $i \in [1,10]$ & $\sqrt{1 + \bar z_i}$
\end{tabular}
\label{tab:1}
\end{ruledtabular}
\end{center}
\end{table}

\subsection{\label{sec:cosmic shear}$3 \times 2$ Point Statistics}
Gravitational lensing of distant galaxies induces non-zero $E$-mode power in the angular correlations between galaxy ellipticities. For tomographic bin pairs $\{ i,j \}$, with $i<j$, the relevant two-point statistic in harmonic space is the shear power spectrum, $C_{i j}^{\gamma \gamma}(\ell)$. Galaxy ellipticites also tidally align with large nearby dark matter halos leading to additional subdominant -- yet important contributions -- to the observed lensing spectrum, $C_{i j}^{\rm LL}(\ell)$. These are referred to as intrinsic alignments. In particular the term, $C_{i j}^{\gamma \mathrm{I}}(\ell)$, accounts for correlation between shear acting on foreground galaxies and intrinsic alignments. This is taken to be zero because a background IA should not be correlated with a foreground shear. The $C_{i j}^{ \mathrm{I} \gamma}(\ell)$ terms gives the correlation between foreground IA and background shear and a $C_{i j}^{ \mathrm{I} \mathrm{I} }(\ell)$ term accounts for the auto-correlation in IA. Finally a shot-noise term $N_{i j}^{\mathrm{LL}}(\ell)$ accounts for the Poisson noise associated with the dispersion in the intrinsic ellipticities of galaxies before being sheared. We are left with
\begin{equation}  \label{eq:LL}
    C_{i j}^{\rm LL}(\ell)=C_{i j}^{\gamma \gamma}(\ell)+C_{i j}^{\mathrm{I} \gamma}(\ell)+C_{i j}^{\mathrm{II}}(\ell) + N_{i j}^{\mathrm{LL}}(\ell).
\end{equation}
\par The clustering of foreground galaxies is correlated with (lensing) structure which shears background galaxies. This gives rise to the galaxy-galaxy lensing signal and we write the observed spectrum as $C_{i j}^{\rm GL}(\ell)$. One must also account for the intrinsic alignment of galaxies so that
\begin{equation} \label{eq:GL}
    C_{i j}^{\rm GL}(\ell)=C_{i j}^{\mathrm{g} \gamma }(\ell)+C_{i j}^{\mathrm{I} \mathrm{g}}(\ell).
\end{equation}
The terms $C_{i j}^{\mathrm{g} \mathrm{I}}(\ell)$ and $C_{i j}^{\gamma \mathrm{g}}(\ell)$ are taken to be zero. There are also no shot-noise contributions since the dispersion in shear and clustering are uncorrelated.
\par Finally the observed clustering spectrum $C_{i j}^{\rm GG}(\ell)$ is given as the sum of the cosmological signal and the shot-noise contributions
\begin{equation} \label{eq:GG}
    C_{i j}^{\rm GG}(\ell)=C_{i j}^{\mathrm{gg}}(\ell) + N_{i j}^{\mathrm{gg}}(\ell). 
\end{equation}
\par In practice we use the $C(\ell)$'s computed in~\cite{blanchard2019euclid} (hereafter EF19 for `{\it Euclid} forecasting'), to which we refer the reader to Sect. 3 for detailed models of the individual terms in equations~(\ref{eq:LL}) - (\ref{eq:GG}). In brief, EF19 assume the Limber\footnote{The Limber relation is invalid for $\ell \lesssim 100$~\cite{Fang:2019xat, kitchinglimits} so that any future study must include the full non-Limber expressions.} ~\citep{loverdelimber}, flat-sky~\citep{kitchinglimits}, Zeldovich ~\citep{kitching2017unequal} and reduced shear approximations~\citep{deshpande2020euclid}. It has also recently been shown that $k$-cut cosmic shear reduces the impact of the reduced shear approximation~\cite{deshpande2020accessing}. For the IA terms we use an extended nonlinear alignment model (eNLA)~\citep{joachimi2011constraints}. The global IA amplitude is written as a product, $C_{\rm IA} A_{\rm IA}$, where $A_{\rm IA}$ is left as a free parameter and $C_{\rm IA}$ is fixed. Two free parameters $\eta_{\rm IA}$ and $\beta_{\rm IA}$ act as power law indices for the redshift and luminosity dependence respectively. The model reduces to the standard nonlinear alignment model~\citep{bridle2007dark} if $\eta_{\rm IA}$ and $\beta_{\rm IA}$ are taken to be zero. We also ignore the impact of magnification bias~\citep{thiele2020disentangling} and redshift-space disortions~\cite{Hamilton:1997zq,Padmanabhan:2006cia}. Finally, it is assumed that the galaxy bias is multiplicative leading to $10$ additional nuisance parameters $b_i$ for each tomographic bin $i$. The fiducial values are taken to be $b_i = \sqrt{1 + \bar z_i}$, where $\bar z_i$ is the mean redshift of tomographic bin $i$ in the absence of photometric redshift errors. A summary of the survey set-up, cosmological parameters, and their fiducial values are given in Tab.~\ref{tab:1}. In all cases we consider all $\ell \in [10, 5000]$, except when explicitly stated otherwise.

\subsection{\label{sec:The Bernardeau-Nishimichi-Taruya (BNT) Transformation}The Bernardeau-Nishimichi-Taruya (BNT) Transformation}

\begin{figure}[!hbt]
\includegraphics[width=1.\linewidth]{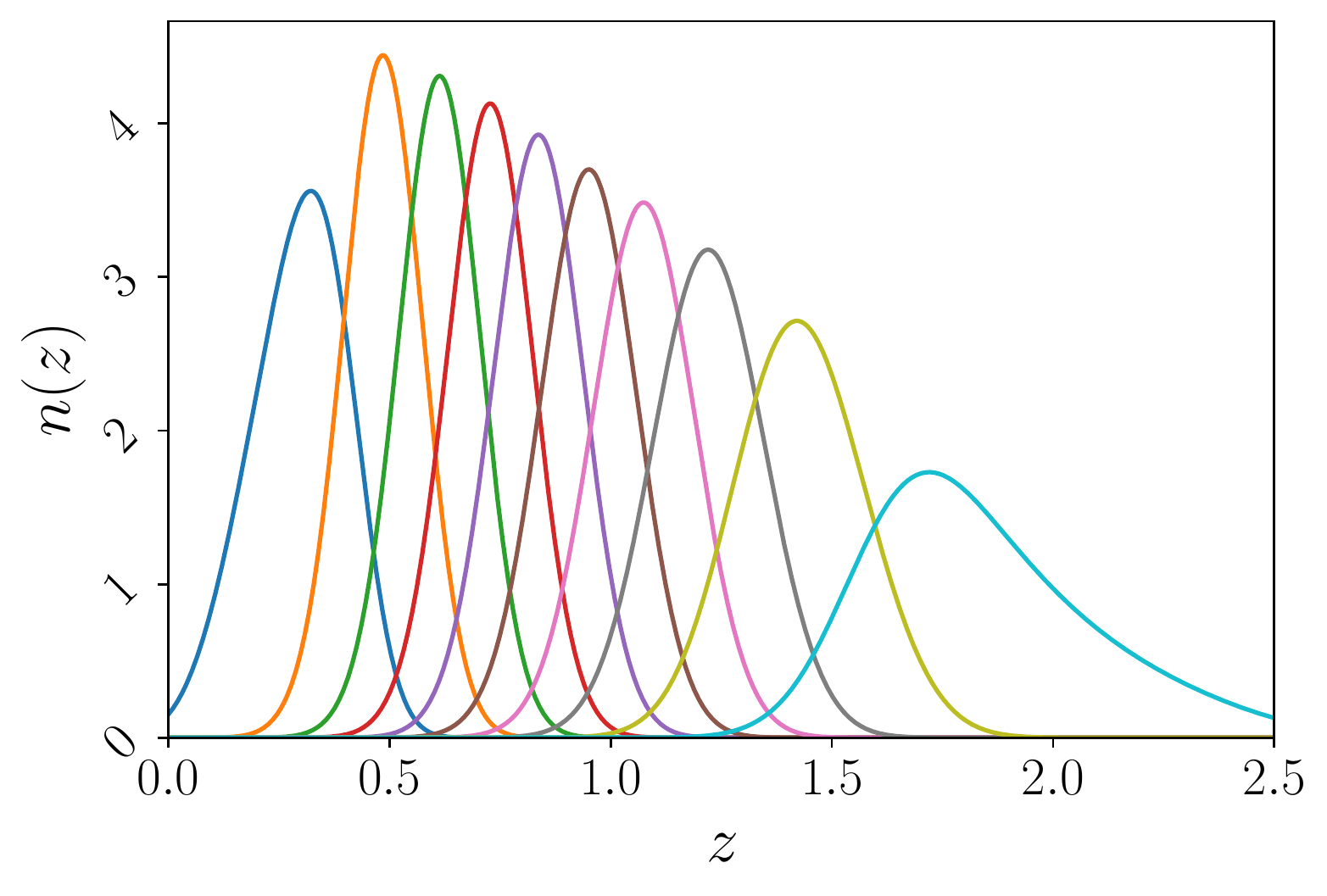}
\includegraphics[width=1.\linewidth]{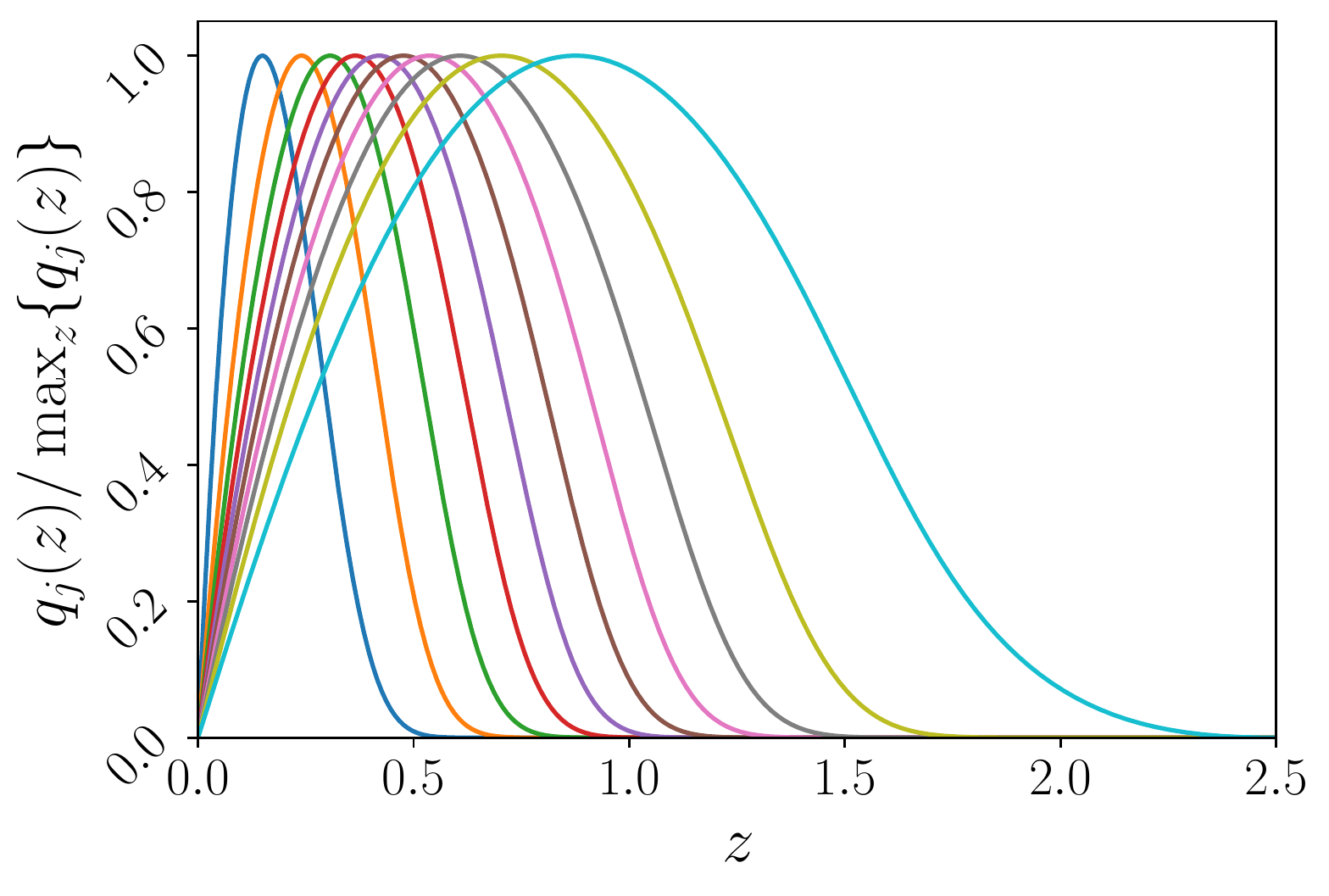}
\includegraphics[width=1.\linewidth]{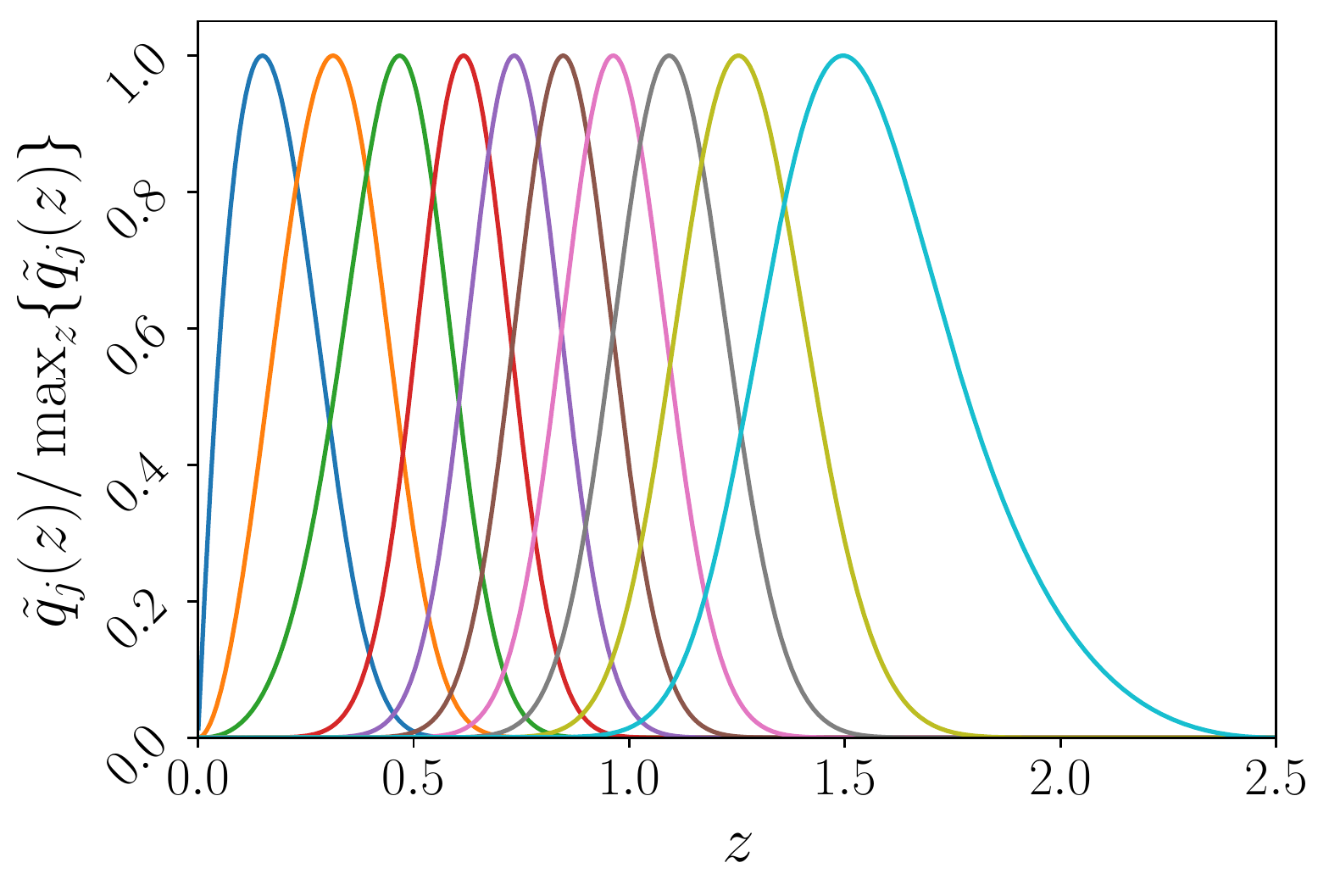}
\caption{{\bf Top}: The radial PDF for the $10$ tomographic bins considered in this work. {\bf Middle}: The corresponding lensing efficiency kernels normalised against there maximum values. These are broad which means that the lensing signal in each tomographic bin is sensitive to structure over a broad range in redshift. {\bf Bottom}: BNT transformed kernels. These are narrow in redshift making it possible to relate physical structure scales, $k$, with angular wavemodes, $\ell$.}
\label{fig:kernels}
\end{figure}

For each tomographic bin, $i$, the lensing efficiency kernel, $q_i(\chi)$, gives the sensitivity of the lensing signal to structure at comoving distance $\chi$. It is defined by
\begin{equation}
q_i(\chi) = \frac{3}{2}\Omega_{\rm m} \left(\frac{H_0}{c}\right)^2 \frac{\chi}{a(\chi)} \int_{\chi}^{\chi_{\rm H}} {\rm d}\chi' n_i(\chi') \frac{\chi'-\chi}{\chi'},
\end{equation}
 where $\chi_{\rm H}$ is the distance to the horizon, $H_0$ is the Hubble parameter, $\Omega_{\rm m}$ is the fractional matter density parameter, $c$ is the speed of light, and $a$ is the scale factor.
 \par As in EF19, we assume that galaxies are equipartitioned into $10$ tomographic bins, and that
\begin{equation}
n \left( z \right)  \propto \left( z/ z_{e} \right) ^2 \exp \left[ {- \left( z / z_e \right) ^ {3/2}} \right],
\end{equation}
with $z_e = 0.9 / \sqrt{2}$, smoothed by the Gaussian kernel
\begin{equation} \label{eq:photo error}
\begin{aligned}
p \left( z' | z \right) &= \frac{0.9}{\sqrt{2 \pi} \sigma(z)} \exp \left[ -\frac{1}{2} \left(\frac{z-z'}{\sigma (z)} \right) ^ 2 \right] \\ 
&+ \frac{0.1}{\sqrt{2 \pi} \sigma(z)} \exp \left[ -\frac{1}{2} \left(\frac{z-z'-0.1}{\sigma (z)} \right) ^ 2 \right] 
\end{aligned}
\end{equation}
where $z'$ is the measured redshift accounting for photometric redshift uncertainty, with $\sigma(z) =0.05 \left(  1 + z \right).$ This functional from is motivated in Sect. 3.1 of~\cite{kitching2008systematic}.
The resulting $n_j(z)$ and $q_j(z)$ are plotted in Fig.~\ref{fig:kernels}.  The lensing efficiency kernels are broad in redshift which implies that the shear signal for galaxies inside each tomographic bin is sensitive to lensing structure over a broad range in redshift.
\par One can define new kernels which are narrow in redshift by taking a linear combination of tomographic bins 
\begin{equation}
\tilde q_i(\chi) = M_{ij}q_j(\chi),
\end{equation}
where $M$ is the Bernardeau-Nishimichi-Taruya (BNT) transform matrix.\footnote{Although the BNT transform formally has some cosmological dependence, it is shown in~\citet{bernardeau2014cosmic, taylor2020x} that this is an extremely small effect in practice. Nevertheless, we compute the BNT transform at the fiducial cosmology used in the rest of the paper.} This transform was proposed in~\citet{bernardeau2014cosmic} and the generalisation to the continuous case is explicitly written down in~\citet{taylor2020x}. 
\par  The BNT matrix, M, is an $N 
\times N$ matrix where $N$ is the number of tomographic bins. After setting $M_{ii} = 1$ for all $i$ and $M_{ij} = 0$ for $i<j$, the remaining BNT matrix elements are found by solving the system
\begin{equation}
\begin{aligned}
\sum_{j=i-2}^{i}M^{ij} &= 0\\
\sum_{j=i-2}^{i}M^{ij}B^j &= 0,
\end{aligned}
\end{equation}
where
\begin{equation}
    B^j = \int_0^{z_{\rm max}} {\rm d}z' \frac{n_j (z')}{\chi(z')}
\end{equation}
and $z_{\rm max}$ is the maximum redshift of the survey.
In this work we compute the BNT matrix, $M$, using the publicly available code at: \hyperlink{https://github.com/pltaylor16/x-cut}{https://github.com/pltaylor16/x-cut}. 
\par The BNT transformed kernels are shown in Fig.~\ref{fig:kernels}. These are narrow implying each new tomographic bin is only sensitive to lensing structure over a small range in redshift. This allows one to more precisely relate angular scale, $\ell$, and physical scale, $k$, which we formalise in the next section.

\subsection{\label{sec:cosmic shear}$3 \times 2$ Point $k$-cut Statistics}
\par One can also make the BNT transformation at the level of the two-point statistics by applying the BNT transformation each time the lensing efficiency kernel appears in the theoretical expressions in the spectra.\footnote{The intrinsic alignment terms have different kernels from the $\gamma \gamma$ term leading to some suboptimality in the transformation. However, IA contributions account for only $\sim 10 \%$ of the signal, so this is a small effect.}
\par In case of the lensing spectrum this is referred to as the $k$-cut cosmic shear~\citep{taylor2018k-cut} spectrum and is given by
\begin{equation}    
\widetilde C^{\rm LL}_{ij} (\ell) = M_{ik} C^{\rm LL}_{kl} (\ell) \left( M^T \right)_{lj}.
\end{equation}
In~\citet{taylor2020x} this was extended to galaxy-galaxy lensing in configuration space. In harmonic space the galaxy-galaxy lensing spectrum, $\widetilde C^{\rm GL}_{ij}$, is given by
\begin{equation}
   \widetilde C^{\rm GL}_{ij} (\ell) =  C^{\rm GL}_{ik} (\ell) \left( M^T \right)_{kj}, 
\end{equation}
The galaxy clustering spectrum is left unchanged so that
\begin{equation}
    \widetilde C^{\rm GG}_{ij} (\ell) = C^{\rm GG}_{ij} (\ell).
\end{equation}
\par Each BNT transformed tomographic bin is only sensitive to structure inside a narrow redshift range. Now one can define a `typical' comoving distance, $\chi_i$, to each comoving bin by taking a weighted average\footnote{To be extremely conservative, one could instead use the lower bound of the kernel, but it was found in~\cite{taylor2018k-cut} that using the mean nearly completely removes sensitivity below the desired cut.} of $\chi$ values over the BNT kernel
\begin{equation}
    \chi^\gamma_i = \frac{\int_0^{\chi_H} {\rm d} \chi \ \chi \tilde q_i(\chi)}{\int_0^{\chi_H} {\rm d} \chi \  \tilde q_i(\chi)}.
\end{equation}
In the case of galaxy clustering the kernels, $n_i(\chi)$, are already narrow and we define the typical distance as
\begin{equation}
    \chi^{\rm G}_i = \frac{\int_0^{\chi_H} {\rm d} \chi \ \chi n_i(\chi)}{\int_0^{\chi_H} {\rm d} \chi \  n_i(\chi)}.
\end{equation}
\par Now using the Limber relation implies that cutting $\ell$-modes with $\ell_i > k \chi_i$, for each tomographic bin, nearly completely removes sensitivity to small-scale structure above some predefined target $k$-mode, $k$. Because we are dealing with two-point statistics, for each tomographic bin pair ($i < j$), there are two relevant kernels and hence -- from the Limber relation -- two choices for the angular scale cut. We take the most conservative of the two cuts and remove
\begin{equation}
\begin{aligned}
\ell_i >& \min \{ k \chi^{\gamma}_i,  k \chi^{\gamma}_j \}, \\
\ell_i >& \min \{ k \chi^{\rm G}_i,  k \chi^{\gamma}_j \}, \\
\ell_i >& \min \{ k \chi^{\rm G}_i,  k \chi^{\rm G}_j \}, \\
\end{aligned}
\end{equation}
for the cosmic shear, galaxy-galaxy lensing, and galaxy clustering cases respectively. If this $\ell$-value is larger than the global $\ell_{\rm max}$, then no cut is made for these combination of bins. We refer to the resulting BNT transformed and cut estimators as $k$-cut $3 \times 2$ point statistics.
\par We note that it is straightforward to extend a traditional $3 \times 2$ point likelihood analysis to $k$-cut $3 \times 2$ statistics. The main obstacle may appear to be the computation of a valid covariance matrix to form the likelihood. However the `likelihood sampling method' defined in~\cite{taylor2020x} can be used to transform the standard $3 \times 2$ covariance into a $k$-cut $3 \times 2$ point covariance in a few CPU minutes. This method works by drawing noise realisations from $\mathcal{N} (0, \widehat C)$, where $\widehat C$ is an estimate of the covariance of $C_{ij}(\ell)$, before BNT-transforming the mock realisations and directly estimating the $k$-cut cosmic shear covariance matrix from the samples.\footnote{At present the likelihood sampling method assumes the likelihood is Gaussian (although a more realistic likelihood could easily be used instead, as required). While the Gaussian likelihood approximation is valid at the level of parameter constraints for cosmic shear alone~\citep{taylor2019cosmic, lin2019non}, this must be explicitly checked for $3 \times 2$ point statistics.} To make a fair comparison with EF19, we do not perform Markov Chain Monte Carlo (MCMC) forecasting, focusing exclusively on Fisher matrix forecasting.

\subsection{\label{sec:Fisher Forecasting} Fisher Forecasting}
We assume a Gaussian covariance neglecting both the Super-Sample Covariance (SSC) and Non-Gaussian (NG) terms, as in EF19.
Defining
\begin{equation} \label{eq:fish1}
    \Delta C_{i j}^{A B}(\ell)=\sqrt{\frac{2}{(2 \ell+1) f_{\mathrm{sky}} \Delta \ell}}C_{i j}^{AB}(\ell),
\end{equation}
where $\Delta \ell$ is the multipole bandwidth, the Fisher matrix for the $3 \times 2$ point statistics using a second-order covariance\footnote{It is shown in~\cite{carron2013assumption} that the the fourth-order covariance and second-order covariance Fisher formalisms will yield the same forecasts.} is given by
\begin{equation} \label{eq:fish2}
\begin{aligned}
F_{\alpha \beta}^{\mathrm{XC}}=\sum_{\ell=\ell_{\min }}^{\ell_{\max }} \sum_{A B C D} \sum_{i j, m n} \frac{\partial C_{i j}^{A B}(\ell)}{\partial \theta_{\alpha}}\left[\Delta C^{-1}(\ell)\right]_{j m}^{A B} \\ \times \frac{\partial C_{m n}^{C D}(\ell)}{\partial \theta_{\beta}}\left[\Delta C^{-1}(\ell)\right]_{n i}^{C D},
\end{aligned}
\end{equation}
where $f_{\rm sky}$ is the fractional sky coverage, $\alpha$ and $\beta$ label the cosmological parameters, $i,j,m$ and $n$ label tomographic bins and $A$, $B$, $C$, and $D$ correspond to either lensing or galaxy clustering. 
\par To make forecasts for the $k$-cut $3 \times 2$ point statistics we make the replacement
\begin{equation}
    C_{ij}^{AB} (\ell) \rightarrow \widetilde C_{ij}^{AB} (\ell)
\end{equation}
in Eqs. (\ref{eq:fish1})--(\ref{eq:fish2}), taking $\ell$-cuts as required.
\par Using the publicly available\footnote{\url{https://github.com/euclidist-forecasting/fisher_for_public}} Fisher matrix for the {\it Euclid} spectroscopic clustering analysis (see~EF19), we can also include information from the spectroscopic survey
\begin{equation} \label{eq:spec fisher}
    F_{\alpha \beta}^{\mathrm{tot}}=F_{\alpha \beta}^{\mathrm{XC}} + F_{\alpha \beta}^{\mathrm{spec}}.
\end{equation}
In this paper we will consider both $F_{\alpha \beta}^{\mathrm{XC}}$ and $F_{\alpha \beta}^{\mathrm{spec}}$. This expression ignores cross-correlations that may exist between the spectroscopic and photometric probes. The majority of the spectroscopic sample lies above $z = 0.9$, so the cross-correlation with the photometric probes is expected to be small. For more details about the spectroscopic Fisher forecasts, we refer the reader to Sect. 3.2 of EF19.
\par In all that follows we use the dark energy Figure of Merit (FOM)~\citep{albrecht2006report} to compare the constraining power for different $k$-cuts. The FOM is proportional to the area enclosed by the $1\sigma$ contours in the $w_0-w_a$ plane. As in~\cite{albrecht2006report,blanchard2019euclid} we define the FOM as
\begin{equation}
    {\rm FOM}_{w_0 w_a} = \sqrt{\widetilde F_{w_0 w_a}},
\end{equation}
where $\widetilde F_{w_0 w_a}$ is the Fisher matrix after marginalising over all the other parameters, which is equivalent to taking the Schur complement~\citep{kitching2009fisher}. 
\par We stress that the results are subject to the modelling assumptions made in Sect.~\ref{sec:cosmic shear}. Additional nuisance parameters and the inclusion of SSC terms in the covariance will all degrade the FOM.

\section{Results} \label{sec:Results}

We use the $C_\ell$s and derivatives computed in EF19. The reader is referred to Sect. 4 of this work for a detailed discussion of the computation of the second derivatives. We perform a quick check to validate that we reproduce the results in EF19, using the standard $3 \times  2$ point statistics before exploring the $k$-cut constraints. 

\subsection{\label{sec:Verification} Verification}
\begin{figure}[!hbt]
\centering
\includegraphics[width=1.0\linewidth]{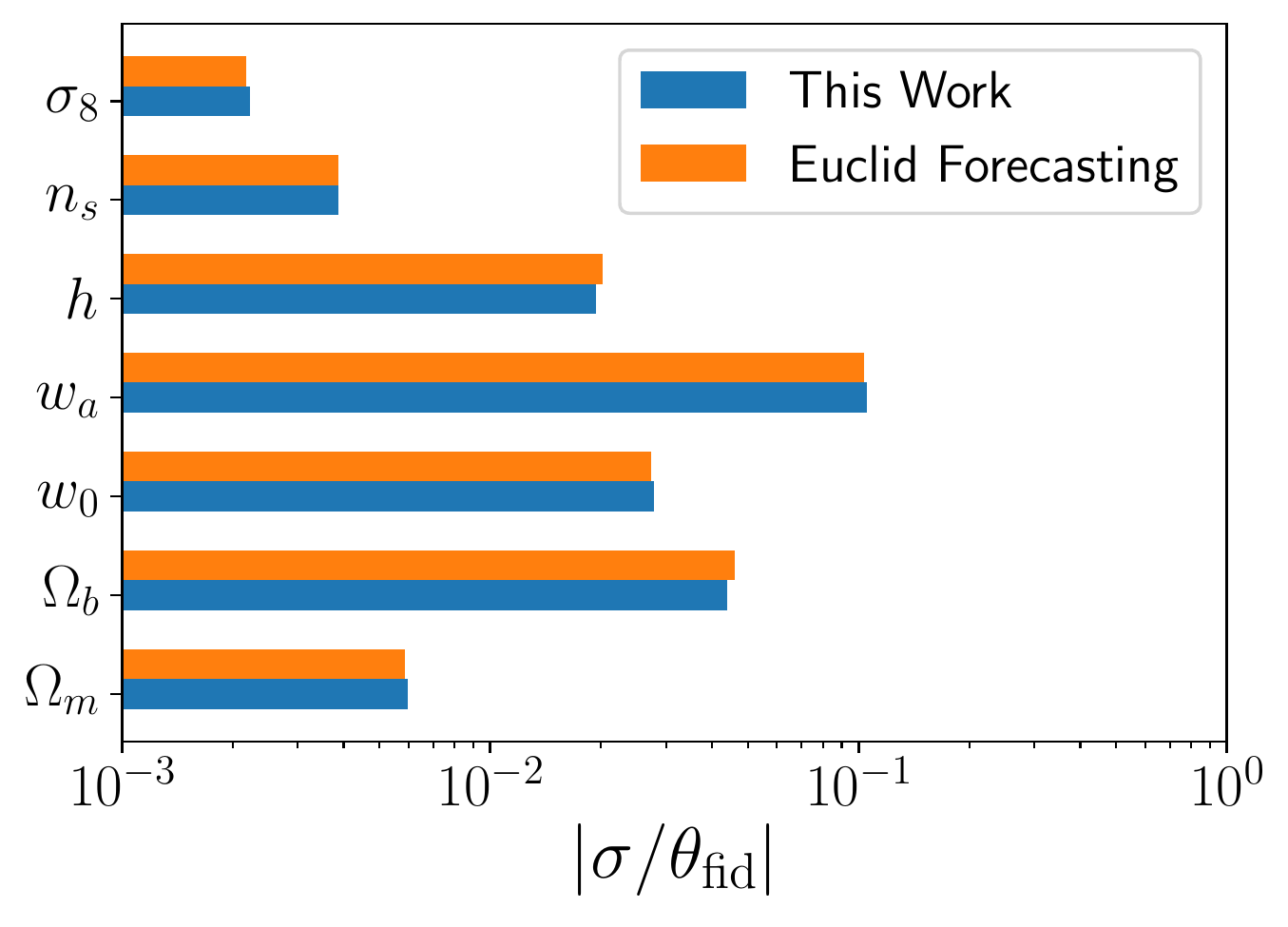}
\caption{The absolute value of the computed marginal errors relative to the fiducial parameter values in EF19 (orange) and this work (blue). We find excellent agreement, validating our Fisher matrix code.}
\label{fig:comparison}
\end{figure}

\par Taking a cut at $\ell = 3000$ for galaxy clustering and galaxy-galaxy lensing while allowing the lensing spectra to range up to $\ell = 5000$, we compute the Fisher matrix for the $3 \times 2$ point statistics. The choice of $\ell$-cuts is `the optimistic case' considered in EF19. After marginalising over the nuisance parameters, we compute the absolute value of the ratio of the marginal error relative to the fiducial values, $|\sigma / \theta_{\rm fid}|$,\footnote{For the parameter $w_a$ the fiducial value is zero, so we use $\sigma(w_a)$ instead of $|\sigma / \theta_{\rm fid}|$.} and compare our results to EF19 in Fig.~\ref{fig:comparison}. We find excellent agreement. The FOM differs by $1 \%$.

\subsection{\label{sec:32 Point Forecasts} Fiducial $3 \times 2$ Point Forecasts}

\begin{figure}[!hbt]
\centering
\includegraphics[width=1.0\linewidth]{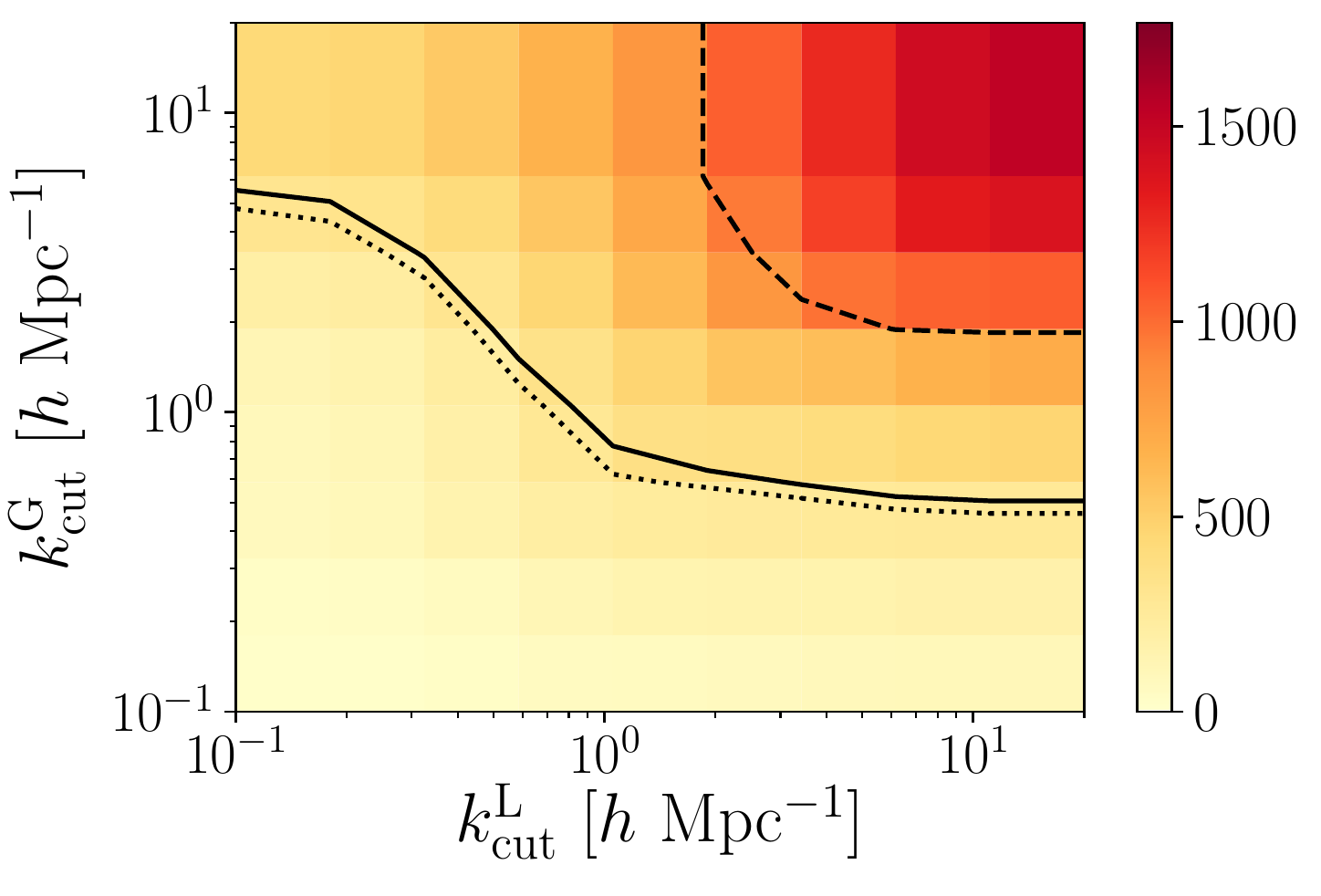}
\includegraphics[width=1.0\linewidth]{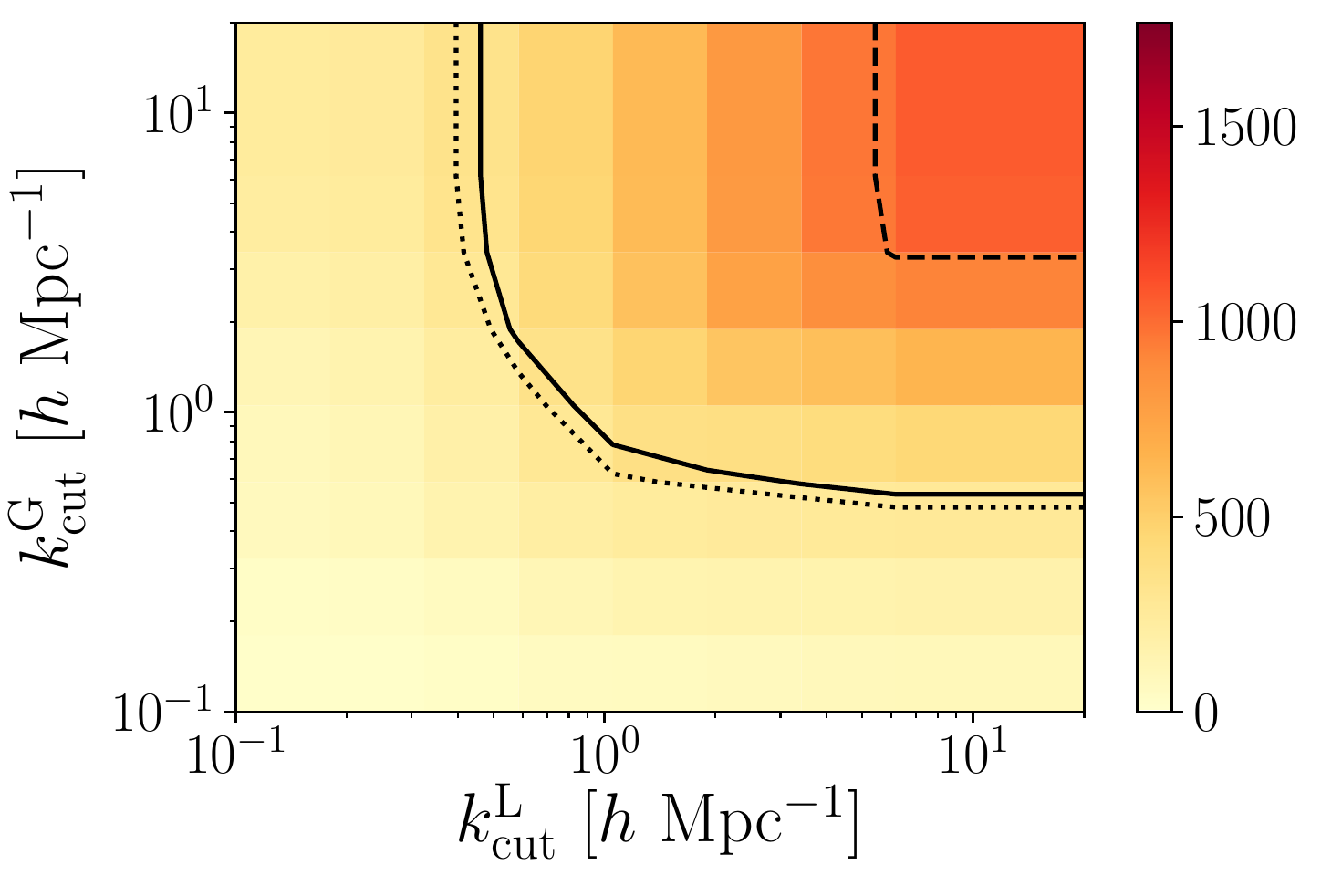}
\caption{Dark energy Figure of Merit (FOM). $k^{\rm L}_{\rm cut}$ gives the $k$-cut scale for cosmic shear while $k^{\rm G}_{\rm cut}$ gives the cut scale for galaxy clustering and galaxy-galaxy lensing. Dotted and dashed continuous black lines correspond to FOMs of 367 and 1033 respectively. These are the FOMs for the `pessimistic' and `optimistic' cases in EF19 which are summarised in Tab.~\ref{tab:2}. The solid black line marks a FOM of 400 from the {\it Euclid} Red Book {\bf Top:} Global $\ell_{\rm max} = 5000.$  A cut scale of $k \sim 2.6 \ h \ {\rm Mpc} ^ {-1}$ yields a similar FOM to the optimistic case in EF19. {\bf Bottom:} Global $\ell_{\rm max} = 3000.$}
\label{fig:FOM_no_spec}
\end{figure}

\begin{figure}[!hbt]
\centering
\includegraphics[width=1.0\linewidth]{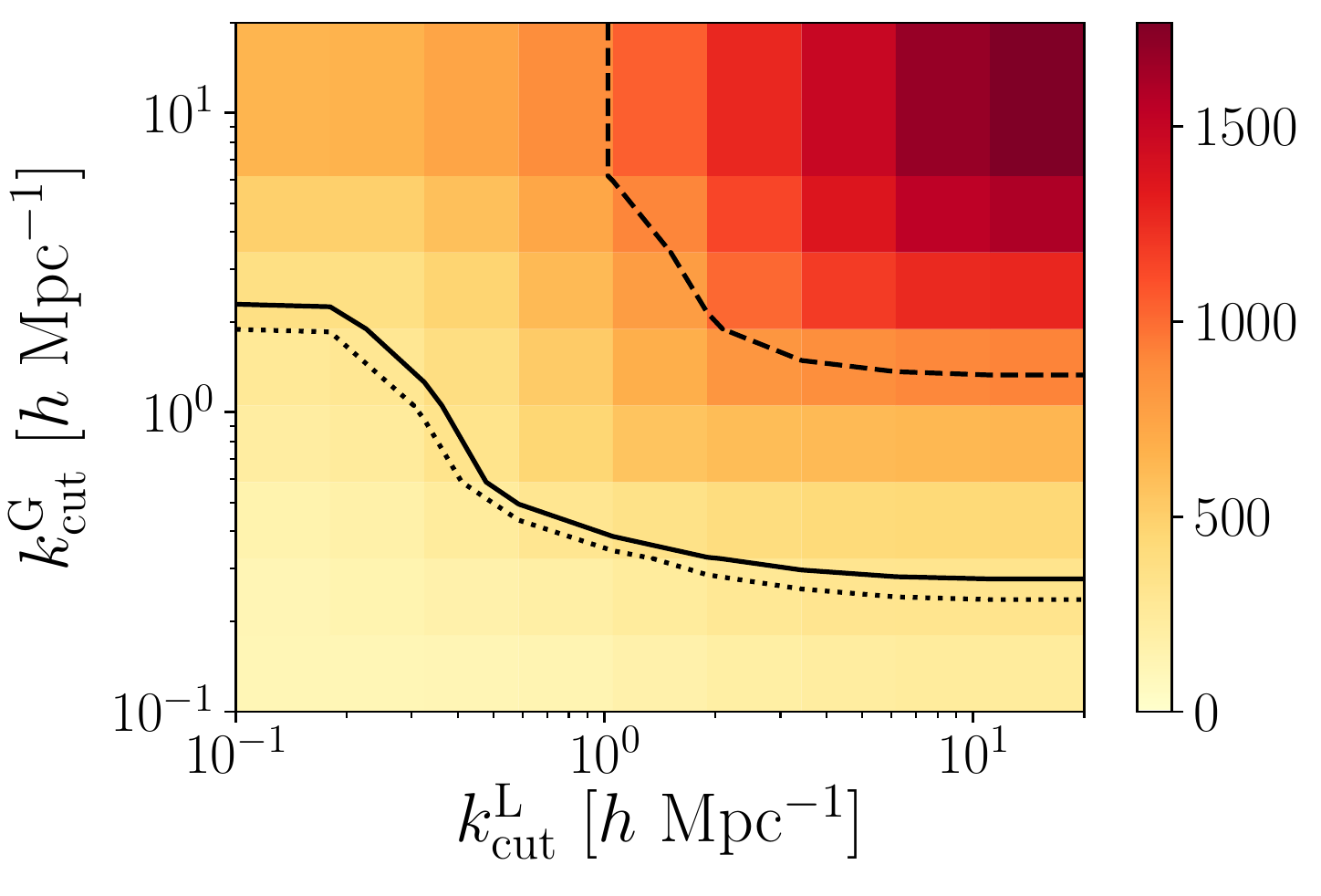}

\caption{Same as Fig.~\ref{fig:FOM_no_spec} except this we include the spectroscopic clustering information by adding the spectroscopic clustering Fisher matrix as in Eq.~(\ref{eq:spec fisher}). For the spectroscopic forecasts we take the `optimistic settings' from EF19 (we refer the reader to Sect. 4 of that work for more details). Compared to the fiducial case, the inclusion of the spectroscopic data increases the FOM by $20 \%$ while using the same cut scales ($k^{\rm L}_{\rm cut} = k^{\rm G}_{\rm cut} =  2.6 \ h \ {\rm Mpc} ^{-1}$ and $\ell_{\rm max} = 5000$).}
\label{fig:FOM_spec}
\end{figure}

\begin{figure}[!hbt]
\centering
\includegraphics[width=1.0\linewidth]{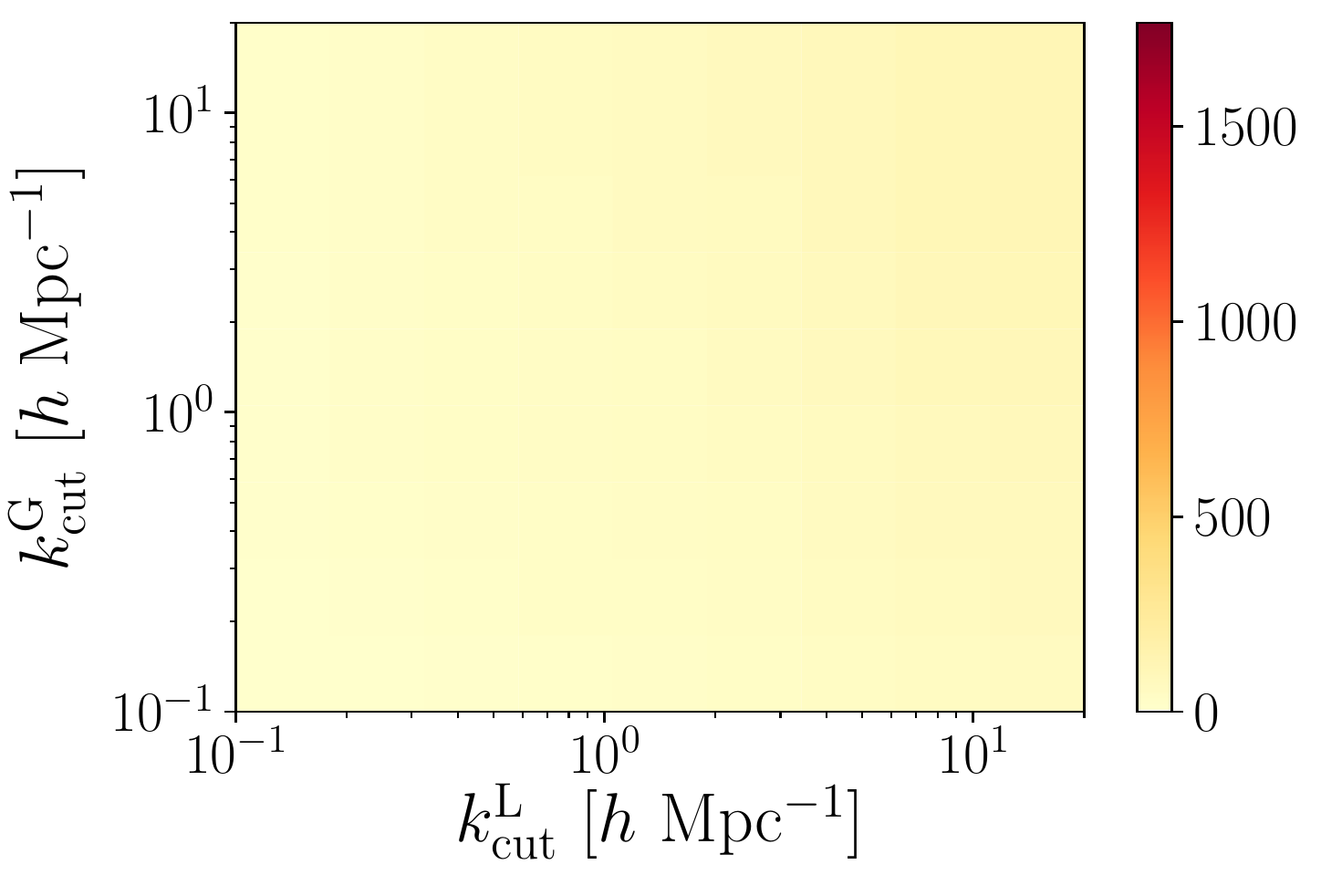}
\includegraphics[width=1.0\linewidth]{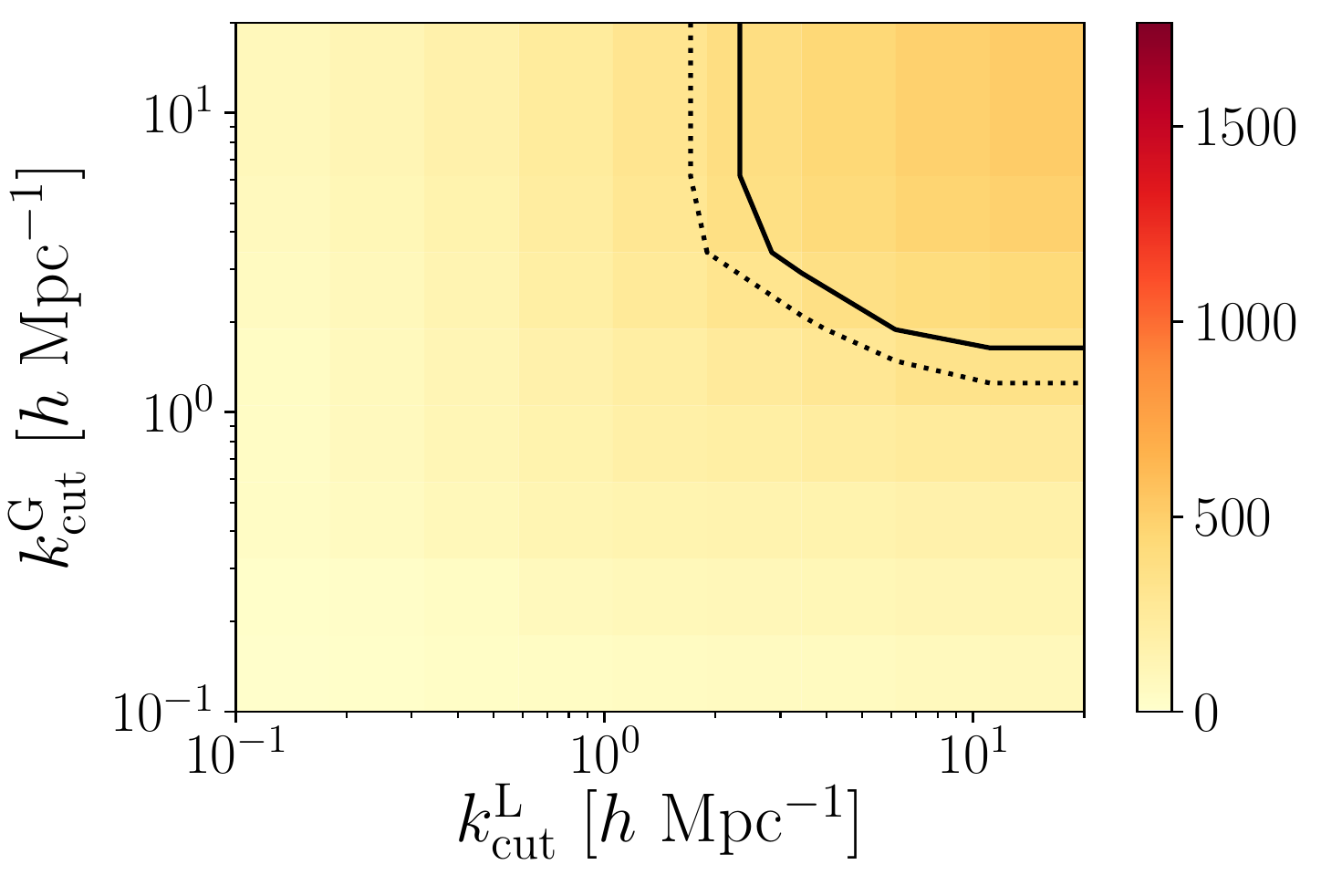}
\includegraphics[width=1.0\linewidth]{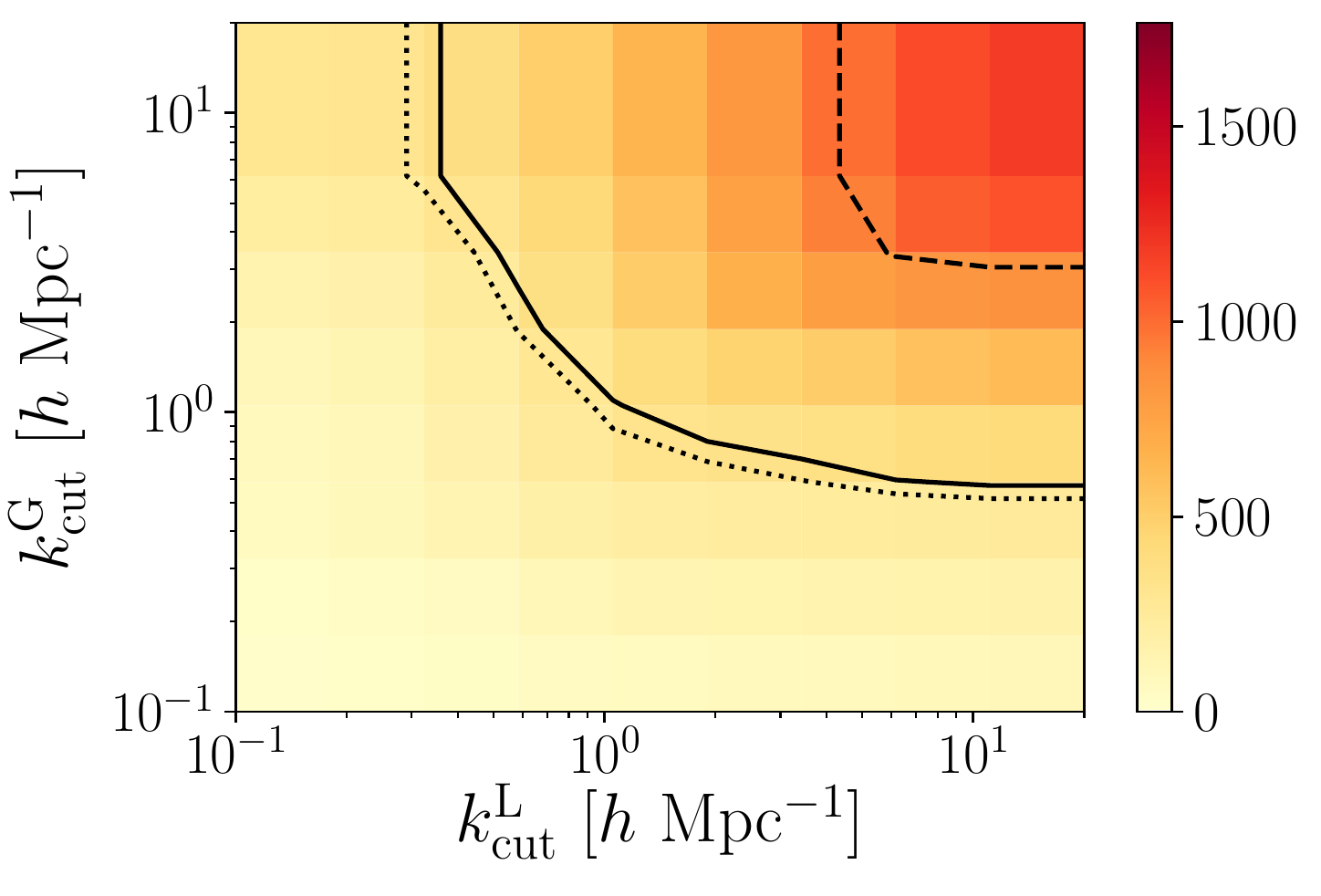}

\caption{Same as Fig.~\ref{fig:FOM_no_spec} but using a sub-sample of the available galaxies for the photometric clustering analysis {\bf Top}: FOM using $1 \%$ of the  available galaxies. {\bf Middle} FOM using $10 \%$ of the  available galaxies. {\bf Bottom}: FOM using $50 \%$ of the  available galaxies. At the fiducial cut scale, $k^{\rm L}_{\rm cut} = k^{\rm G}_{\rm cut} =  2.6 \ h \ {\rm Mpc} ^{-1}$, the FOMs for a subsample of $1 \%$, $10 \%$, $50 \%$ and $100 \%$ of available galaxies are 73, 378, 820, and 1018 respectively.}
\label{fig:FOM_sub}
\end{figure}

We examine the change in the FOM for different $k$-cuts in Fig.~\ref{fig:FOM_no_spec}. Even after taking $k$-cuts one may still need to take an $\ell$-cut to remove detector systematics so we consider both $\ell_{\rm max} = 5000$ (top) and  $\ell_{\rm max} = 3000$ (bottom), before taking additional $\ell$-cuts to make the $k$-cut. The colour scale indicates the FOM. On the axes, $k^{\rm L}_{\rm cut}$ indicates the $k$-mode cut scale for cosmic shear while
$k^{\rm G}_{\rm cut}$ gives the cut scale for galaxy clustering and galaxy-galaxy lensing.\footnote{We choose to have the same cut scale for galaxy-galaxy lensing and clustering since they both have dependence on the galaxy bias. In a more realistic setting, this is uncertain at high-$k$.}
The solid black line corresponds to the FOM target of $400$ from the {\it Euclid} Red Book~\citep{laureijs2010euclid}. {\it It should be noted that the Red Book forecasts are for a non-flat cosmology, so the results presented here are not strictly comparable.} The dotted and dashed continuous lines indicate FOMs of 367 and 1033, respectively. These are the FOMs for the `pessimistic' and `optimistic' cases in EF19 which are summarised in Tab.~\ref{tab:2}.

\begin{table}[hbt!]
\caption{Overview of the cut scales for the `optimistic' and `pessimistic' analyses in EF19 and the fiducial and conservative $k$-cut $3 \times 2$ point analyses used in this work (see Sect.~\ref{sec:32 Point Forecasts}).}
\label{table:params}
\begin{center}
\begin{ruledtabular}
\begin{tabular}{ lccccccc }
  {} & Optimistic & Pessimistic & Fiducial & Conservative\\
  \hline
  \hline
  $\ell^{\rm G}_{\rm cut}$  & 5000 & 1500 & 5000 & 5000\\
   $\ell^{\rm L}_{\rm cut} $ & 3000 & 750 & 5000 & 5000\\
  $k^{\rm G}_{\rm cut} [h \ {\rm Mpc} ^{-1}]$ & N/A & N/A & 2.6 & 0.4 \\
  $k^{\rm L}_{\rm cut}[h \ {\rm Mpc} ^{-1}]$ &  N/A & N/A & 2.6 & 1.0\\
  FOM & 1033 & 367 & 1018 & 283 \\
\end{tabular}
\label{tab:2}
\end{ruledtabular}
\end{center}
\end{table}

\par For the case $\ell_{\rm max} = 5000$, a cut of $k \sim 2.6 \ \mathnormal{h} \ {\rm Mpc} ^{-1}$ for clustering, lensing, and cross-correlations gives a similar FOM to the optimistic case in EF19, while $k \sim 0.7 \ h \ {\rm Mpc} ^{-1}$ yields a FOM of $400$ from the {\it Euclid} Red Book.
\par Modelling uncertainties are problematic at high $k$ but other systematics (e.g. point-spread function corrections) become a problem at high $\ell$~\citep{paykari2020euclid}. For this reason we also consider the case where $\ell_{\rm max} = 3000$. Then a cut scale of $k \sim 4 \ h \ {\rm Mpc} ^{-1}$ and $k \sim 0.7 \ h \ {\rm Mpc} ^{-1}$ for both clustering and lensing are needed to match the optimistic and Red Book FOMs respectively.
\par We take as our fiducial case $\ell_{\rm max} = 5000$ with $k^{\rm L}_{\rm cut} = k^{\rm G}_{\rm cut} =  2.6 \ h \ {\rm Mpc} ^{-1}$, because it has a FOM of 1018, close to the optimistic case in EF19. We also consider a conservative $k$-cut case with $\ell_{\rm max} = 5000$ with $k^{\rm L}_{\rm cut} = 1$ and $k^{\rm G}_{\rm cut} =  0.4 \ h \ {\rm Mpc} ^{-1}$ to reflect current galaxy bias and baryonic physics modelling limitations. In this case the FOM is 283.

\subsection{\label{sec:Inclusion of Spectroscopic Data} Inclusion of Spectroscopic Clustering}
In Fig.~\ref{fig:FOM_spec} we again plot the FOM as function of cut scales ($k^{\rm G}_{\rm cut}$ and $k^{\rm L}_{\rm cut}$) but this time we also include information by adding the spectroscopic clustering Fisher matrix as in Eq.~(\ref{eq:spec fisher}). For the spectroscopic Fisher matrix, we use the optimistic spec-$z$ settings in EF19 (the reader is referred to Sect. 4 of this work for more details). 
\par Including the information from spectroscopic clustering analysis means that it is possible to take a cut at a smaller $k$-value while achieving the same FOM. For example a FOM of 400 meeting the Red Book requirements can be achieved by taking a $k$-cut at $0.6 \ h \ {\rm Mpc} ^{-1}$. The conservative scale cut case ($k^{\rm L}_{\rm cut} = 1$ and $k^{\rm G}_{\rm cut} =  0.4 \ h \ {\rm Mpc} ^{-1}$) also meets the Red Book requirements with FOM of 416. Meanwhile at the fiducial cut scale of $k^{\rm L}_{\rm cut} = k^{\rm G}_{\rm cut} = 2.6 \ h \ {\rm Mpc} ^{-1}$, the inclusion of spectroscopic information improves the FOM by $19 \%$.

\subsection{\label{sec:reduced tracer population} Reduced Tracer Population}
\par So far we have assumed that $100 \%$ of the available galaxies are used in the photometric clustering analysis. However current Stage III $3 \times 2$ point analyses~\citep{abbott2018dark, heymans2020kids} use only a fraction of the galaxies for the clustering analysis compared to the cosmic shear measurement. This simplifies the analysis as galaxy bias is strongly dependent on type and using bright galaxies minimises the impact of foregrounds~\citep{elvin2018dark}. In this section we explore the impact of only using sub-sample of the available galaxies in the photometric clustering analysis. Specifically we recompute the FOM after multiplying the galaxy-clustering shot-noise term, defined in Eq.~(\ref{eq:GG}), by $1/F$, where $F$ is the fraction of galaxies used in the photometric clustering analysis.   
\par The results of this computation are shown in Fig.~\ref{fig:FOM_sub} which are worth comparing to Fig.~\ref{fig:FOM_no_spec}. The top, middle and bottom subplots correspond to using $1 \%$, $10 \%$ and $50 \%$ of the available galaxies respectively. 
\par When only $1 \%$ of the galaxies are used, the FOM never exceeds 400, while for $10 \%$, the FOM never exceeds 1000 -- for any choice of $k$-cut. When $50 \%$ of the galaxies are used, we achieve the `optimistic' case FOM described in EF19 when we take a cut at $k \sim 5 \ h \ {\rm Mpc} ^{-1}$ and a FOM of $400$ with a cut at $k \sim 1 \ h \ {\rm Mpc} ^{-1}$.
\par At the fiducial cut scale, $k^{\rm L}_{\rm cut} = k^{\rm G}_{\rm cut} =  2.6 \ h \ {\rm Mpc} ^{-1}$, the FOMs for a subsample of $1 \%$, $10 \%$, $50 \%$, and $100 \%$ of available galaxies are 73, 378, 820, and 1018. Thus increasing the subsample from $10 \%$ to $50 \%$ more than doubles the FOM while expanding the subsample from $50 \%$ to $100 \%$ increases the FOM by $24 \%$. This gain is similar to including the spectroscopic clustering (see the previous section) in the analysis. It is evident that including a larger fraction of the available galaxies in the photometric clustering analysis is one of the primary drivers of the FOM in {\it Euclid}, provided that we are able to model the small scales down to $k \sim 2.6 \ h \ {\rm Mpc} ^{-1}$.
\par On the other hand if the analysis is restricted to our conservative choice of scale cuts, $k^{\rm L}_{\rm cut} = 1$ and $k^{\rm G}_{\rm cut} =  0.4 \ h \ {\rm Mpc} ^{-1}$, the affects of a reduced tracer population are not as dramatic. In this case the FOMs for a subsample of $1 \%$, $10 \%$, $50 \%$, and $100 \%$ of available galaxies are 41, 161, 256, and 282. The reason that the FOM is less dependent on the fraction of galaxies used when we take a more conservative scale cut is because we are then less sensitive to high $\ell$ modes where shot-noise is larger relative to the signal.
\par Meanwhile when we include the spectroscopic information as in Sect.~\ref{sec:Inclusion of Spectroscopic Data} taking the fiducial cut scale of $k^{\rm L}_{\rm cut} = k^{\rm G}_{\rm cut} =  2.6 \ h \ {\rm Mpc} ^{-1}$, the FOMs for a subsample of $1 \%$, $10 \%$, $50 \%$, and $100 \%$ of available galaxies are 228, 567, 1008 and 1207. When $50\%$ of the galaxies are used we achieve the `optimistic' case FOM of 1033 for a cut at $k \sim 3 \ h \ {\rm Mpc} ^{-1}$ and a FOM of 400 when we take a cut at $k \sim 0.6 \ h \ {\rm Mpc} ^{-1}$. This is in comparison to the case where we use $100\%$ of the galaxies when we achieve we achieve the `optimistic' case FOM of 1033 for a cut at $k \sim 1 \ h \ {\rm Mpc} ^{-1}$.
\par It should be noted that we have made three useful first-order approximations in this section: 
\begin{itemize}
\item The shape of redshift distribution function $n(z)$ is fixed. In reality each tracer population has its own distribution function changing the global $n(z)$ as more galaxies are included.
\item The photometric uncertainty is fixed. In fact photo-z estimates for the commonly-used LRG subsample are more precise than for most other populations~\citep{rozo2016redmagic}. For this reason our results likely overestimate the information loss from excluding galaxies. 
\item We have also assumed simplistic linear galaxy bias model with only one free parameter per redshift bin. The systematic uncertainty is therefor likely underestimated.
\end{itemize}
Studying the impact of these effects is left to a future work.

%\newpage
\section{Conclusions} \label{sec:Conslusions}
In this paper we have developed the formalism for $k$-cut $3 \times 2$ point statistics and provided Fisher forecasts for {\it Euclid}.  In a more realistic setting one would likely need to include free parameters for multiplicative biases, as well as more complicated models for IA and galaxy bias. One would also need to consider the impact of non-Gaussian~\citep{barreira2018accurate} and super-sample corrections~\citep{hu2003sample} to the covariance. Since the $3 \times 2$ point statistics are not linear in the cosmological parameters, MCMC forecasting would give more realistic constraints. These extensions are left to a future work. 
\par The $k$-cut method efficiently removes sensitivity to small physical scales which are difficult to model. This enables the extraction of useful information at small angular scales which would otherwise need to be completely removed from the analysis. We find that taking a cut at $k = 2.6 \ h \ {\rm Mpc} ^ {-1}$ (while taking a global $\ell_{\rm max} = 5000$) for both galaxy clustering and lensing yields FOM of 1018 which is similar to the `optimistic case' $(\ell_{\rm max} = 5000$ for lensing and  $\ell_{\rm max} = 3000$ for clustering and galaxy-galaxy lensing) in EF19 where a FOM of 1033 is achieved. The final choice of $k$-cut in {\it Euclid} depends on the accuracy of the matter power spectrum model at the time the data arrives. This is left for investigation in a future work.
\par To avoid bias from `observational' systematics (caused by e.g. point-spread function residuals, blending, foreground and charge transfer inefficiency) in $k$-cut $3 \times 2$ point analyses, it may be necessary to take additional angular scale cuts. A thorough investigation of `observational' systematics~\citep{paykari2020euclid} at these typically excluded angular scales (high $\ell$) is warranted. 
\par The clustering part of Stage III $3 \times 2$ point analyses have worked with LRGs~\citep{abbott2018dark} or directly with data from external spectroscopic surveys~\citep{heymans2020kids, van2018kids+, joudaki2018kids} for the clustering analysis. Hence we have investigated the degradation in FOM when only sub population of the available galaxies are used in the clustering analysis. We find this to be one of the primary drivers of the FOM in {\it Euclid}, particularly if we are able to model the observables to small scales.
\par We have demonstrated that $k$-cut $3 \times 2$ point statistics are a viable method to reduce sensitivity to small poorly modelled scales in {\it Euclid}. This comes at virtually no cost given the small computational overhead and the fact that this technique can be used in combination with other mitigation strategies (e.g. marginalising over baryonic feedback nuisance parameters). In light of ever-improving models of small-scale physics, we leave the determination of the optimal cut scale for {\it Euclid}, which must strike a balance between minimising bias and precision, to a future work. Meanwhile we have shown the importance of including as many galaxies in the photometric clustering sample as possible.

\begin{acknowledgements}
 The authors would like to thank Shahab Joudaki for carefully reviewing an earlier version of the paper. We thank the two anonymous referees whose comments have significantly improved the manuscript. PLT acknowledges support for this work from a NASA Postdoctoral Program Fellowship. Part of the research was carried out at the Jet Propulsion Laboratory, California Institute of Technology, under a contract with the National Aeronautics and Space Administration. TDK acknowledges funding from the European Union's Horizon 2020 research and innovation programme under grant agreement No. 776247. ACD acknowledges funding from the Royal Society. The authors acknowledge support from NASA ROSES grant 12-EUCLID12-0004. AP is a UK Research and Innovation Future Leaders Fellow, grant MR/S016066/1.
\AckECol
 
\end{acknowledgements}

\bibliographystyle{aa}
\bibliography{ref}
-------------------------------------------------------------------------\\
%% please do not edit the affiliation list -- contact ECEB Bureau for changes
\noindent $^{1}$ Jet Propulsion Laboratory, California Institute of Technology, 4800 Oak Grove Drive, Pasadena, CA, 91109, USA\\
$^{2}$ Mullard Space Science Laboratory, University College London, Holmbury St Mary, Dorking, Surrey RH5 6NT, UK\\
$^{3}$ INAF-Osservatorio Astronomico di Roma, Via Frascati 33, I-00078 Monteporzio Catone, Italy\\
$^{4}$ Institut de Physique Th\'eorique, CEA, CNRS, Universit\'e Paris-Saclay F-91191 Gif-sur-Yvette Cedex, France\\
$^{5}$ Institut d'Astrophysique de Paris, 98bis Boulevard Arago, F-75014, Paris, France\\
$^{6}$ Institute of Space Sciences (ICE, CSIC), Campus UAB, Carrer de Can Magrans, s/n, 08193 Barcelona, Spain\\
$^{7}$ Institut d’Estudis Espacials de Catalunya (IEEC), Carrer Gran Capit\'a 2-4, 08034 Barcelona, Spain\\
$^{8}$ School of Physics and Astronomy, Queen Mary University of London, Mile End Road, London E1 4NS, UK\\
$^{9}$ INFN-Sezione di Torino, Via P. Giuria 1, I-10125 Torino, Italy\\
$^{10}$ Dipartimento di Fisica, Universit\'a degli Studi di Torino, Via P. Giuria 1, I-10125 Torino, Italy\\
$^{11}$ INAF-IASF Milano, Via Alfonso Corti 12, I-20133 Milano, Italy\\
$^{12}$ AIM, CEA, CNRS, Universit\'{e} Paris-Saclay, Universit\'{e} Paris Diderot, Sorbonne Paris Cit\'{e}, F-91191 Gif-sur-Yvette, France\\
$^{13}$ Instituto de F\'isica T\'eorica UAM-CSIC, Campus de Cantoblanco, E-28049 Madrid, Spain\\
$^{14}$ Universit\'e St Joseph; UR EGFEM, Faculty of Sciences, Beirut, Lebanon\\
$^{15}$ Institut de Recherche en Astrophysique et Plan\'etologie (IRAP), Universit\'e de Toulouse, CNRS, UPS, CNES, 14 Av. Edouard Belin, F-31400 Toulouse, France\\
$^{16}$ Departamento de F\'isica, FCFM, Universidad de Chile, Blanco Encalada 2008, Santiago, Chile\\
$^{17}$ Astrophysics Research Institute, Liverpool John Moores University, 146 Brownlow Hill, Liverpool L3 5RF, UK\\
$^{18}$ INAF-Osservatorio di Astrofisica e Scienza dello Spazio di Bologna, Via Piero Gobetti 93/3, I-40129 Bologna, Italy\\
$^{19}$ INAF-Osservatorio Astronomico di Padova, Via dell'Osservatorio 5, I-35122 Padova, Italy\\
$^{20}$ Max Planck Institute for Extraterrestrial Physics, Giessenbachstr. 1, D-85748 Garching, Germany\\
$^{21}$ INAF-Osservatorio Astrofisico di Torino, Via Osservatorio 20, I-10025 Pino Torinese (TO), Italy\\
$^{22}$ Universit\'e de Paris, CNRS, Astroparticule et Cosmologie, F-75006 Paris, France\\
$^{23}$ INFN-Sezione di Roma Tre, Via della Vasca Navale 84, I-00146, Roma, Italy\\
$^{24}$ Department of Mathematics and Physics, Roma Tre University, Via della Vasca Navale 84, I-00146 Rome, Italy\\
$^{25}$ INAF-Osservatorio Astronomico di Capodimonte, Via Moiariello 16, I-80131 Napoli, Italy\\
$^{26}$ Institut de F\'{i}sica d’Altes Energies (IFAE), The Barcelona Institute of Science and Technology, Campus UAB, 08193 Bellaterra (Barcelona), Spain\\
$^{27}$ Department of Physics "E. Pancini", University Federico II, Via Cinthia 6, I-80126, Napoli, Italy\\
$^{28}$ INFN section of Naples, Via Cinthia 6, I-80126, Napoli, Italy\\
$^{29}$ INAF-Osservatorio Astrofisico di Arcetri, Largo E. Fermi 5, I-50125, Firenze, Italy\\
$^{30}$ Dipartimento di Fisica e Astronomia, Universit\'a di Bologna, Via Gobetti 93/2, I-40129 Bologna, Italy\\
$^{31}$ Centre National d'Etudes Spatiales, Toulouse, France\\
$^{32}$ Institute for Astronomy, University of Edinburgh, Royal Observatory, Blackford Hill, Edinburgh EH9 3HJ, UK\\
$^{33}$ European Space Agency/ESRIN, Largo Galileo Galilei 1, 00044 Frascati, Roma, Italy\\
$^{34}$ ESAC/ESA, Camino Bajo del Castillo, s/n., Urb. Villafranca del Castillo, 28692 Villanueva de la Ca\~nada, Madrid, Spain\\
$^{35}$ INFN-Sezione di Bologna, Viale Berti Pichat 6/2, I-40127 Bologna, Italy\\
$^{36}$ Dipartimento di Fisica "Aldo Pontremoli", Universit\'a degli Studi di Milano, Via Celoria 16, I-20133 Milano, Italy\\
$^{37}$ INFN-Sezione di Milano, Via Celoria 16, I-20133 Milano, Italy\\
$^{38}$ Institute of Theoretical Astrophysics, University of Oslo, P.O. Box 1029 Blindern, N-0315 Oslo, Norway\\
$^{39}$ von Hoerner \& Sulger GmbH, Schlo{\ss}Platz 8, D-68723 Schwetzingen, Germany\\
$^{40}$ Max-Planck-Institut f\"ur Astronomie, K\"onigstuhl 17, D-69117 Heidelberg, Germany\\
$^{41}$ Aix-Marseille Univ, CNRS/IN2P3, CPPM, Marseille, France\\
$^{42}$ Universit\'e de Gen\`eve, D\'epartement de Physique Th\'eorique and Centre for Astroparticle Physics, 24 quai Ernest-Ansermet, CH-1211 Gen\`eve 4, Switzerland\\
$^{43}$ Department of Physics and Helsinki Institute of Physics, Gustaf H\"allstr\"omin katu 2, 00014 University of Helsinki, Finland\\
$^{44}$ NOVA optical infrared instrumentation group at ASTRON, Oude Hoogeveensedijk 4, 7991PD, Dwingeloo, The Netherlands\\
$^{45}$ Argelander-Institut f\"ur Astronomie, Universit\"at Bonn, Auf dem H\"ugel 71, 53121 Bonn, Germany\\
$^{46}$ Institute for Computational Cosmology, Department of Physics, Durham University, South Road, Durham, DH1 3LE, UK\\
$^{47}$ Istituto Nazionale di Astrofisica (INAF) - Osservatorio di Astrofisica e Scienza dello Spazio (OAS), Via Gobetti 93/3, I-40127 Bologna, Italy\\
$^{48}$ Universit\'e de Paris, F-75013, Paris, France, LERMA, Observatoire de Paris, PSL Research University, CNRS, Sorbonne Universit\'e, F-75014 Paris, France\\
$^{49}$ California institute of Technology, 1200 E California Blvd, Pasadena, CA 91125, USA\\
$^{50}$ Observatoire de Sauverny, Ecole Polytechnique F\'ed\'erale de Lau- sanne, CH-1290 Versoix, Switzerland\\
$^{51}$ European Space Agency/ESTEC, Keplerlaan 1, 2201 AZ Noordwijk, The Netherlands\\
$^{52}$ Department of Astronomy, University of Geneva, ch. d'\'Ecogia 16, CH-1290 Versoix, Switzerland\\
$^{53}$ INAF-Osservatorio Astronomico di Trieste, Via G. B. Tiepolo 11, I-34131 Trieste, Italy\\
$^{54}$ Department of Physics and Astronomy, University of Aarhus, Ny Munkegade 120, DK–8000 Aarhus C, Denmark\\
$^{55}$ Perimeter Institute for Theoretical Physics, Waterloo, Ontario N2L 2Y5, Canada\\
$^{56}$ Department of Physics and Astronomy, University of Waterloo, Waterloo, Ontario N2L 3G1, Canada\\
$^{57}$ Centre for Astrophysics, University of Waterloo, Waterloo, Ontario N2L 3G1, Canada\\
$^{58}$ Space Science Data Center, Italian Space Agency, via del Politecnico snc, 00133 Roma, Italy\\
$^{59}$ Institute of Space Science, Bucharest, Ro-077125, Romania\\
$^{60}$ Universit\"ats-Sternwarte M\"unchen, Fakult\"at f\"ur Physik, Ludwig-Maximilians-Universit\"at M\"unchen, Scheinerstrasse 1, 81679 M\"unchen, Germany\\
$^{61}$ INFN-Padova, Via Marzolo 8, I-35131 Padova, Italy\\
$^{62}$ Dipartimento di Fisica e Astronomia “G.Galilei", Universit\'a di Padova, Via Marzolo 8, I-35131 Padova, Italy\\
$^{63}$ Centro de Investigaciones Energ\'eticas, Medioambientales y Tecnol\'ogicas (CIEMAT), Avenida Complutense 40, 28040 Madrid, Spain\\
$^{64}$ Infrared Processing and Analysis Center, California Institute of Technology, Pasadena, CA 91125, USA\\
$^{65}$ Instituto de Astrof\'isica e Ci\^encias do Espa\c{c}o, Faculdade de Ci\^encias, Universidade de Lisboa, Tapada da Ajuda, PT-1349-018 Lisboa, Portugal\\
$^{66}$ Departamento de F\'isica, Faculdade de Ci\^encias, Universidade de Lisboa, Edif\'icio C8, Campo Grande, PT1749-016 Lisboa, Portugal\\
$^{67}$ Universidad Polit\'ecnica de Cartagena, Departamento de Electr\'onica y Tecnolog\'ia de Computadoras, 30202 Cartagena, Spain\\
$^{68}$ Kapteyn Astronomical Institute, University of Groningen, PO Box 800, 9700 AV Groningen, The Netherlands\\

%\bsp
%\label{lastpage}

\end{document}